\documentclass[aps,prl,twocolumn,reprint,groupedaddress,floatfix,showpacs,footinbib]{revtex4-1}

\usepackage[pdftex]{graphicx}
\usepackage{amsmath,amssymb}
\usepackage{bm}
\usepackage{color}
\usepackage[normalem]{ulem}
\usepackage[usenames,dvipsnames]{xcolor}
\usepackage{eufrak}
\usepackage{bbm}
\usepackage{pdfpages}

\newcommand{\will}[1]{{\color{black}{{#1}}}}
\newcommand{\willB}[1]{{\color{black}{{#1}}}}
\newcommand{\tr}[1]{{\color{black}{{#1}}}}

\newcommand{\tm}[1]{{\color{black}{{#1}}}}

\newcommand{\willC}[1]{{\color{black}{{#1}}}}

\begin{document}

\title{{Guiding catalytically active particles with chemically patterned surfaces}}

\author{W. E. Uspal}
\email[Corresponding author: ]{uspal@is.mpg.de}
\author{M. N. Popescu}
\author{S. Dietrich}
\author{M. Tasinkevych}
\affiliation{Max-Planck-Institut f\"{u}r Intelligente Systeme, Heisenbergstr. 3, 
D-70569 Stuttgart, Germany}
\affiliation{IV. Institut f\"ur Theoretische Physik, Universit{\"a}t Stuttgart, 
Pfaffenwaldring 57, D-70569 Stuttgart, Germany}

\date{\today}

\begin{abstract}
Catalytically active Janus particles suspended in solution create gradients in the chemical 
composition of the solution along their surfaces, as well as along any nearby 
container walls. The former leads to self-phoresis, while the latter gives rise to 
chemi-osmosis, providing an additional contribution to self-motility. 
Chemi-osmosis strongly depends on the molecular interactions between the 
diffusing chemical species and the wall. We show analytically, using an approximate 
``point-particle'' approach, that by chemically patterning a planar substrate  one can direct the motion of Janus particles: the induced chemi-osmotic flows can cause particles to either ``dock'' 
at the chemical step between the two  materials, or to follow a chemical 
stripe. These theoretical predictions are confirmed by full numerical calculations. 
Generically, docking occurs for particles which tend to move away from their 
catalytic caps, while stripe-following occurs in the opposite case. Our analysis 
reveals the physical mechanisms governing this behavior.\newline 
\end{abstract}


\maketitle

\label{Intro}
 
The endowment of micrometer sized objects with elements of complex, life-like behavior issuing from simple and controllable physico-chemical components and 
forces is a challenging step towards the development of far-reaching potential applications. The active particles developed in the last decade 
\cite{paxton06,ebbens10} can ``swim'' within a liquid environment, as well as sense and respond to (according 
to their design) local conditions or fields (e.g., surfaces or hydrodynamic flow \cite{palacci15,uspal15b}). These features indeed evoke primitive aspects of cellular life.

Catalytic Janus particles activate, over a fraction of their 
surface, chemical reactions in the surrounding solution. The resulting gradients in 
chemical composition along the surface of an individual 
particle, in combination with the interaction between the molecules of the 
solution and the particle, drive directed motion via, e.g., 
self-diffusiophoresis (for electrically neutral molecules) 
\cite{golestanian05,golestanian07,howse07,baraban12,poon13} or self-electrophoresis (for charged species) 
\cite{paxton04,ebbens14,brown14}.

For mechanical swimmers (e.g., bacteria) under confinement, rigid (soft) boundaries
provide a generic no-slip (continuous shear stress) boundary condition for the 
solvent velocity \cite{lauga08,spagnolie12,lopez14}. For catalytic 
Janus particles, however, boundaries additionally affect the distribution of chemicals 
along the surface of the particle, and thus the self-phoretic motion 
\cite{popescu09,crowdy13,uspal15,ibrahim15}. Furthermore, chemical gradients can drive 
surface flows along the confining boundaries, i.e., the ``dual'' phenomenon of 
chemi-osmosis occurs \cite{anderson89,derjaguin:47}. The chemi-osmotic flows extend 
into the solution and couple back to the particle (see, e.g., the ``chemi-osmotic 
surfers'' discussed in Refs.  \cite{palacci13, palacci14, palacci15}.) Therefore, 
the motility of catalytic Janus particles near a rigid, impenetrable boundary has, in 
general, contributions from both self-diffusiophoresis and chemi-osmosis.

Recently, it has been \willB{shown} that a solid wall with a 
spatially varying slip length can direct the motion of a mechano-elastic 
model of \textit{E. coli} \cite{hu15}. \willC{That patterning regulates how the surface \textit{passively} reflects the  flows created by the swimmer.} In the case of a chemical 
microswimmer near a wall, the particle induces a local \willC{chemi-osmotic surface flow, i.e., an \textit{active} hydrodynamic response.} \willC{The strength of this surface flow is governed by the so-called surface mobility, which is a material dependent parameter.} This raises the issue of
whether a self-induced locking to directed motion can 
occur if the wall is patterned \will{with different materials}. Here, we derive analytical expressions for the contribution of chemi-osmotic flow to the particle velocity based upon a multipolar description 
of the chemical activity of the particle. These expressions exhibit
excellent agreement with the results of detailed numerical calculations. 
We find that spherical particles designed such that they move \textit{towards} 
their catalytic caps can follow a chemical stripe, while particles 
which move \textit{away} from their caps can dock at a chemical step
between two substrate materials. The physical mechanisms driving these behaviors are identified.

\label{Model}

\begin{figure}[!t]
\includegraphics[width=8.cm]{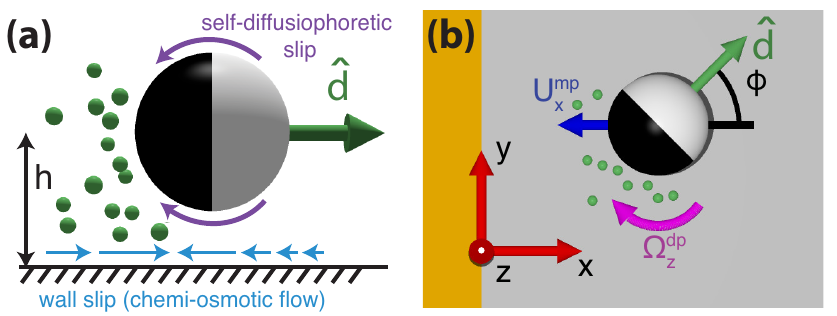}
\caption{\label{fig:figure_1} (a) A catalytic Janus particle with radius $R$
above a planar wall at height $h$. The cap of the particle (black) produces solute molecules 
(green spheres). The effective interaction of the solute and the particle 
surface drives a surface flow (purple). If the effective interaction is 
repulsive (attractive), the flow is towards (away from) the cap, leading to 
``inert-forward'' (``catalyst-forward'') self-diffusiophoretic motion in the $\bf{\hat{d}}$ (-$\bf{\hat{d}}$) direction. 
The solute also drives chemi-osmotic flow on the wall (blue; the direction shown 
is for a repulsive interaction between solute and wall). 
(b) A particle above a patterned substrate. The gray region is more repulsive to the solute than 
the orange one. Consequently, chemi-osmotic flow drives translation (blue arrow) 
and rotation (magenta arrow) of the particle. The abbreviations \textit{mp} and 
\textit{dp} stand for monopole and dipole, respectively.
}
\end{figure}

\textit{Model.}-- We consider a spherical particle of radius $R$ with axisymmetric 
catalyst coverage (Fig.~\ref{fig:figure_1}). We foresee that two features of the 
activity will be important: the particle is a net producer of solute, and production 
is localized to a subregion of the particle surface. Formally, within a multipole 
expansion for the solute  field, these two aspects correspond to a point 
source of solute (a \textit{monopole}) and a \textit{dipolar} pair of a 
source and a sink, for which we anticipate the following roles. A point source 
located above a planar substrate produces a rotationally symmetric solute 
distribution; hence, if the substrate is patterned, the monopole drives \textit{translation} of 
the particle \textit{in the direction defined by the pattern}.  A dipole intrinsically 
has a direction; therefore, it drives \textit{rotation} of the particle relative to the 
patterned-defined direction.

We assume a stationary reference frame in which the instantaneous position of the 
particle is $\mathbf{x}_{0}=(x_{p},y_{p},h)$. The orientation 
$\bf{\hat{d}}$ of the particle is directed along the axis of symmetry from the 
catalytic region to the particle center (Fig. \ref{fig:figure_1}). The particle 
emits solute at a rate (areal density per time) $\alpha(\theta')$ over its 
surface, where the latitudinal angle $\theta'$ is defined with respect to 
$\bf{\hat{d}}$. If the P\'{e}clet number $Pe\,{\equiv}\,U_{0}R/D\ll{1}$, where $U_{0}$ is a 
characteristic particle velocity and $D$ is the diffusion coefficient of the solute 
molecule, the solute number density field $c(\mathbf{x})$ is approximately 
quasi-static, i.e., it obeys $\nabla^{2}c=0$, with a boundary condition 
$-D\,\hat{\mathbf{n}} \cdot \nabla c=\alpha(\theta')$ on the particle surface, with the
normal $\hat{\mathbf{n}}$ pointing towards the liquid.  The impenetrable planar wall at 
$z=0$ imposes the condition $\hat{\mathbf{n}} \cdot ^{•}\nabla{c}=0$ on $c(\mathbf{x})$. \will{We shall develop an analytical framework valid for arbitrary $\alpha(\theta')$.} 
\will{We choose to specifically consider a hemispherical cap which emits solute at a constant 
rate per area $\kappa$, such that $\alpha(\theta') = \kappa$ over the cap and $\alpha(\theta') = 0$ over the inert particle face. }

We employ the classical theory of neutral diffusiophoresis to describe particle motion \cite{anderson89}. The interaction of the solute molecules with a bounding surface drives surface flows which
are modeled with an effective slip boundary condition $\mathbf{v}_{s}(\mathbf{x}_{s})=-b(\mathbf{x}_{s})\nabla_{||}c$, where 
$\nabla_{||}\equiv(\mathbbm{1}-\mathbf{\hat{n}}\mathbf{\hat{n}})\cdot\nabla$ and 
$\mathbf{x}_{s}$ is a location on the surface. The material dependent parameter $b(\mathbf{x}_{s})$ encapsulates the details of the 
interaction  \footnote{\will{As an example, consider an effective surface potential of the solute in the form of a square well of range $L$ and depth $-W$ (in units of the thermal energy). In this case $b =  L^2 [\exp(W)-1]$ (see Eq. (11) in Ref. \cite{anderson89}). Note that a variation of the range of the potential well by a factor of 1.8 is sufficient to change the value of $b$ by a factor of 3; similar changes in $b$ would be produced by an increase in the potential well depth from 0.5 to 1.}}. The surface flows drive flow in the bulk solution. We assume small Reynolds numbers $Re\equiv\rho{U}_{0}R/\eta$, 
where $\rho$ and $\eta$ are the mass density and the viscosity, respectively, 
of the solution \footnote{\textcolor{black}{For a particle of radius $R \sim 1\;{\mu m}$ propelled by a self-generated oxygen gradient through water with $U_{0} \sim 5\;{\mu m/s}$, one has $Re \approx 10^{-5}$ and $Pe \approx 10^{-3}$ \cite{uspal15b}.}}. The bulk fluid velocity $\mathbf{u}(\mathbf{x})$ and the 
pressure $P(\mathbf{x})$ obey the Stokes equation $-\nabla{P}+\eta\nabla^{2}\mathbf{u}=0$ and incompressibility $\nabla\cdot\mathbf{u}=0$. The velocity boundary conditions are  \tm{$\mathbf{u}\lvert_{wall}=\mathbf{v}_{s}(\mathbf{x}_{s})$} on the wall, and \tm{$\mathbf{u}\lvert_{part}=\mathbf{U}+\bm{\Omega}\times(\mathbf{x}_{s}-\mathbf{x}_{0})+\mathbf{v}_{s}(\mathbf{x}_{s})$} on the particle \tm{surfaces, respectively}. $\mathbf{U}$ and $\bm{\Omega}$ are \tm{unknown} translational and angular velocities of the particle, respectively, which are determined by imposing that the particle is force and torque free. The boundary conditions include \will{activity}-induced flows
at the wall (chemi-osmosis) and the particle (self-diffusiophoresis.) The linearity 
of the Stokes equation allows these contributions to $\mathbf{U}$ and $\bm{\Omega}$ to be calculated separately and superposed, 
i.e., we may write $\mathbf{U}=\mathbf{U}^{ws}+\mathbf{U}^{sd}$ and $\bm{\Omega}=\bm{\Omega}^{ws}+\bm{\Omega}^{sd}$, where the superscripts indicate \textit{w}all \textit{s}lip and \textit{s}elf-\textit{d}iffusiophoresis. 

In the following, we restrict $\mathbf{\hat{d}}$ to the $x-y$ plane ($\hat{\mathbf{d}} \cdot \hat{\mathbf{z}} = 0$) and take $h$ to be constant. \will{This simplifying assumption of quasi-2D motion allows us to focus on the basic features of the particle behavior which can be obtained from surface patterning. It can be imposed externally, e.g., by using magnetic fields and particles containing a magnetic core \cite{baraban12b}. Moreover, as discussed in the conclusions, quasi-2D motion is spontaneously realized by particles with certain surface chemistries or non-spherical shapes.} \tm{The effect of an inert uniform wall on $\mathbf{U}^{sd}$ and $\bm{\Omega}^{sd}$ has been studied in detail in Ref. \cite{uspal15}, where it was shown that $\mathbf{U}^{sd}$ depends only on $h$ and $\mathbf{\hat{d}} \cdot \hat{z}$.} \will{Therefore, in the present study we take $\mathbf{U}^{sd} = U^{sd} \hat{\mathbf{d}}$, with $U^{sd}$ treated as an input parameter.}
We recall that for $U^{sd}>0$ ($U^{sd}<0$) the particle moves away from (towards) its cap when it is in the bulk fluid, due to the repulsive (attractive) interaction between the particle and the solute \cite{anderson89}. We restrict our consideration to materials for which ``surfing'' near a uniform substrate  does not change this inert-forward or catalyst-forward character of the motion (the exception is a special case discussed in Sec.~IV.A in the Supplemental Material \footnote{See Supplemental Material for details of the derivations and additional supporting figures.}). \will{We note that $\Omega_{z}^{sd} = 0$ by symmetry, and the assumption of in-plane motion makes ${\Omega}_{x}^{sd}$ and ${\Omega}_{y}^{sd}$ irrelevant here.}

\tm{The problem for $\mathbf{U}^{ws}$ and $\bm{\Omega}^{ws}$ is obtained 
by setting $\mathbf{v}_{s}(\mathbf{x}_{s})=0$ at the particle surface, and employing the Lorentz 
reciprocal theorem \cite{happel65}, which relates the fluid stresses 
$(\bm{\sigma},\bm{\sigma}')$ and velocity fields $(\mathbf{u},\mathbf{u}')$ of two 
solutions for the Stokes equation which share the same geometry. We take our 
``unprimed'' problem to be the one specified above for the six unknowns 
$\mathbf{U}^{ws}$ and $\bm{\Omega}^{ws}$, requiring six ``primed'' subproblems of our choice. 
The interested reader is referred to Sec.~I in the SM for technical details \cite{Note3}.} Numerically, we use the boundary element method (BEM), as detailed in Ref. \cite{uspal15}, to determine $c(\mathbf{x})$ and the six dual solutions \tm{$(\mathbf{u}'^{(j)}, \bm{\sigma}'^{(j)})$, $j=1,..,6$, corresponding to an inactive particle subject to an external force or an external torque along  $\hat{x}$, $\hat{y}$, or $\hat{z}$.} \tm{We obtain analytical expressions after making the following approximations:} \will{(i) We consider only the monopolar and dipolar contributions of the activity to the solute field, and therefore to the particle velocity. Distinguishing these contributions, we write $\mathbf{U}^{ws} \approx \mathbf{U}^{mp} + \mathbf{U}^{dp}$ and $\bm{\Omega}^{ws} \approx \bm{\Omega}^{mp} + \bm{\Omega}^{dp}$. Note that} \tm{for the activity $\alpha(\theta')$ specified above, the monopole strength is $\alpha_{0}=\kappa/2$ and the dipole 
strength is $|\alpha_{1}|=3\kappa/4$ \cite{michelin14}.} \tm{(ii) The effect of the wall on $c(\mathbf{x})$ is accounted for via an image monopole and image dipole at $\mathbf{x}_{I}=(x_{p},y_{p},-h)$. (iii) For the six primed subproblems, we use the image solutions for a point force or torque above a wall \cite{goldman67,blake74}.}

\begin{figure*}[th]
\includegraphics[width=16.cm]{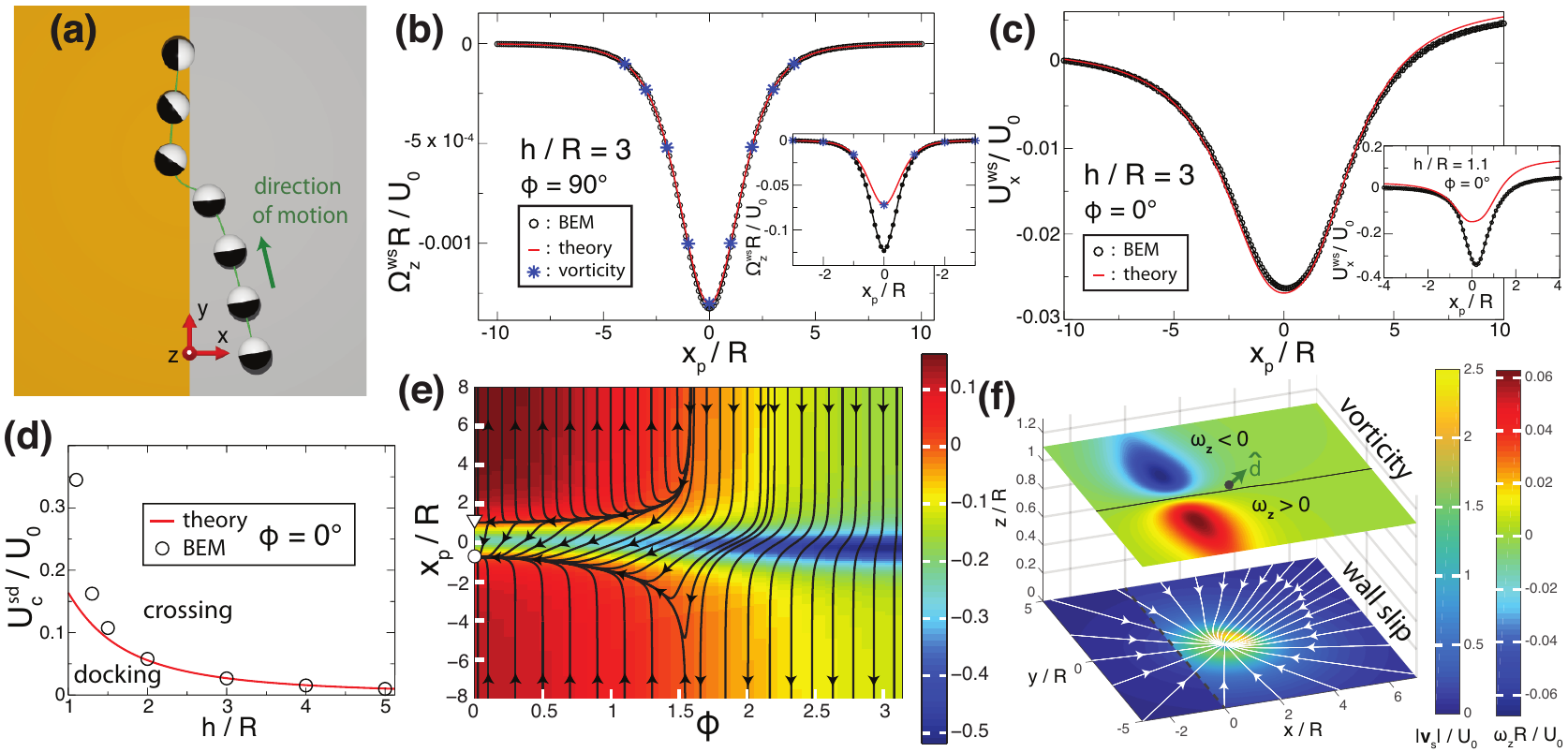}
\caption{\label{fig:figure_2} 
(a) A docking trajectory of an inert-forward particle (with input parameter $U^{sd}/U_{0}=0.1$)
with $h/R=1.1$ near a chemical step with $b_{w}^{r}/b_{w}^{l}=4$, $b_{w}^{l}<0$, 
and $U_{0}\equiv2|b_{w}^{l}|\alpha_{0}/D$, calculated with the BEM. (b) Angular velocity of a particle oriented parallel ($\phi=90^{\circ}$) 
to the step depicted in (a) as a function of $x_{p}$. Black circles were obtained with the BEM, and the red curve stems from Eq. (\ref{eq:line_interface_omegaZ}). \will{Blue stars indicate $\frac{1}{2} \omega_{z}$ at the particle position, where $\bm{\omega}$ is the vorticity, calculated within the ``point-particle'' approximation.} The main panel was calculated for  $h/R=3$ and the inset for $h/R=1.1$. (c) Chemi-osmotic contribution $U_x^{ws}$ to $U_{x}$ 
as a function of $x_{p}$ for a particle with $\phi=0^{\circ}$ (theory corresponds to Eq.~(\ref{eq:line_interface_uxmp}) and  Eq.~(34) in the SM). 
(d) Upper critical $U^{sd}_{c}$ for which a particle can dock at the step depicted in (a) as a function of $h/R$. The red curve was 
calculated using Eq. (\ref{eq:critical_usd}). The symbols were obtained with the 
BEM. (e) Phase plane calculated with the BEM for the system in (a). 
There is an attractor (white circle) at $\phi=0^{\circ}$ and $x_{p}/R=-0.69$ and an unstable fixed point (white triangle). The background color encodes 
$U_{x}/U_{0}$ with $U_x$ obtained from Eq. (\ref{eq:dyn_sys}). The trajectories are mirror symmetric about $\phi=180^{\circ}$, and therefore we omit the region $\phi>180^{\circ}$. \will{(f)  A point-like particle with $\phi = 45^{\circ}$ near the wall as shown in (a). The particle drives a chemi-osmotic flow on the wall with a characteristic monopole plus dipole structure of the streamlines. The surface flow creates a vorticity $\bm{\omega}$ in the bulk fluid. At the position $h/R = 1.1$ and $x_{p}/R = 2$ of the particle, one has $\omega_{z} < 0$, leading to rotation of the particle.}}
\end{figure*}

\textit{Chemical step.}-- \tm{We now consider a substrate with a \tr{chemical step} 
between two materials, such that $b(\mathbf{x}_{s})$ is $b_{w}^{l}$ for $x < 0$ 
and $b_{w}^{r}$ for $x > 0$}. We find (see Sec.~IV.B in the SM \cite{Note3}):
\begin{equation}
\label{eq:line_interface_uxmp}
U_{x}^{mp} = \frac{3hR^{2} \alpha_{0} }{16D} \,
\frac{(b_{w}^{r} - b_{w}^{l}) (h^{2} + 2x_{p}^{2})}{(h^{2} + x_{p}^{2})^{5/2}}.
\end{equation}
By symmetry, one has $U_{y}^{mp}=0$ and $\Omega_{z}^{mp}=0$.  The dipolar contribution can rotate the particle (Fig.~\ref{fig:figure_1}(b)):
\begin{equation}
\label{eq:line_interface_omegaZ}
\Omega_{z}^{dp} = -\frac{3 h R^{3} |\alpha_{1}|}{64 D} 
\frac{(b_{w}^{l}-b_{w}^{r})}{(h^{2} + x_{p}^{2})^{5/2}} \sin(\phi).
\end{equation} 
The lengthy expressions for $\mathbf{U}^{dp}$ are given in the SM \cite{Note3}. In Figs. \ref{fig:figure_2}(b) and (c) we compare the 
predictions of Eqs. (\ref{eq:line_interface_uxmp}) and 
(\ref{eq:line_interface_omegaZ}) with BEM calculations. For $h/R=3$, the 
agreement is excellent; closer to the wall, quantitative differences occur, yet the main trends in the BEM data are captured. This provides an \textit{a posteriori} check that the approximations 
(i)-(iii) are reliable.  

\will{To understand the physical meaning of $U_{x}^{mp}$ and $\Omega_{z}^{dp}$, we examine the flow on the patterned substrate. In  Fig. \ref{fig:figure_2}(f) we show the solution obtained for a point-like particle (i.e., after making the approximations (i)-(iii)); in Fig. 2(a) in the SM, we show the ``exact'' solution, obtained within BEM \cite{Note3}. Clearly, these solutions are approximately identical. Secondly, the streamlines of the surface flow have a monopole plus dipole structure. Interestingly, this structure is independent of the substrate pattern, since it is unaffected by locally rescaling the magnitude of the surface flow velocity  $|\mathbf{v}_{s}(\mathbf{x}_{s})|$ (compare the flow on a uniform substrate in Fig. 2(b) of the SM \cite{Note3}). For a point-like particle, we can numerically calculate the vorticity $\bm{\omega} \equiv \nabla \times \mathbf{u}$ in the bulk created by the surface flow (Fig. \ref{fig:figure_2}(f)). The angular velocity of a tracer particle in a flow field $\mathbf{u}$ is $\bm{\Omega} = \frac{1}{2} \bm{\omega}$. Likewise, we find that $\Omega_{z}^{dp} = \frac{1}{2} \omega_{z}$ at the position of the particle (blue stars in Fig. \ref{fig:figure_2}(b)). This confirms that our analytical expressions treat the particle as a point-like object that locally excites a chemi-osmotic flow and is advected by it as a passive tracer.} 

The trajectory of the particle is obtained by numerical integration of the system
of equations (note that $\Omega_z^{sd}=0$)
\begin{equation}
\label{eq:dyn_sys}
\dot{x}_{p} = U^{sd} \cos(\phi) + U_{x}^{ws}(x_{p}, \phi), \; \; 
\dot{\phi} = \Omega_{z}^{ws}(x_{p}, \phi).
\end{equation}
We find that inert-forward particles ($U^{sd}>0$) can dock at the chemical 
step (Fig.~\ref{fig:figure_2}(a)). As an example, a phase plane showing the 
evolution of $\phi$ and $x_{p}$ for any initial condition, calculated within 
BEM, is given in Fig.~\ref{fig:figure_2}(e). Remarkably, the analytical 
expressions reproduce almost quantitatively this phase plane 
structure (\will{Fig.~3} in the SM \cite{Note3}). The mechanism for docking is as follows. 
$\Omega_{z}^{dp}$ rotates the particle towards $\phi=0^{\circ}$, 
so that the (black) cap faces the region of weaker repulsion (orange left in Fig.~\ref{fig:figure_2}(a)).
 Along $\hat{x}$, the monopole drives the particle away from the step, while
self-diffusiophoresis drives the particle towards the step.
 We estimate that stable docking occurs if the particle cannot cross the step, i.e., if $U_x^{tot}\equiv U^{sd}_{x}+U_{x0}^{mp}+U_{x0}^{dp}\lesssim0$, 
where $U_{x0}^{mp}$ and $U_{x0}^{dp}$ are the monopolar and dipolar contributions 
to $U_{x}^{ws}$ at $x_{p}=0$. The threshold condition $U_x^{tot}=0$ 
predicts the phase boundary $U^{sd}_{c}$ in the $(U^{sd},h)$ plane separating 
docking and crossing (see Sec. IV.B in the SM \cite{Note3}):
\begin{equation}
\label{eq:critical_usd}
U^{sd}_{c} =  \frac{3 \alpha_{0} R^{2} |b_{w}^{l} - b_{w}^{r}|}{16 D h^{2}} + 
\frac{|\alpha_{1}| R^{3} (b_{w}^{r} + b_{w}^{l})}{32 D h^{3}}.
\end{equation}
This expression shows good quantitative agreement with the BEM calculations down to small distances from the wall  (Fig. \ref{fig:figure_2}(d)). 

\begin{figure}[!t]
\includegraphics[width=8.25cm]{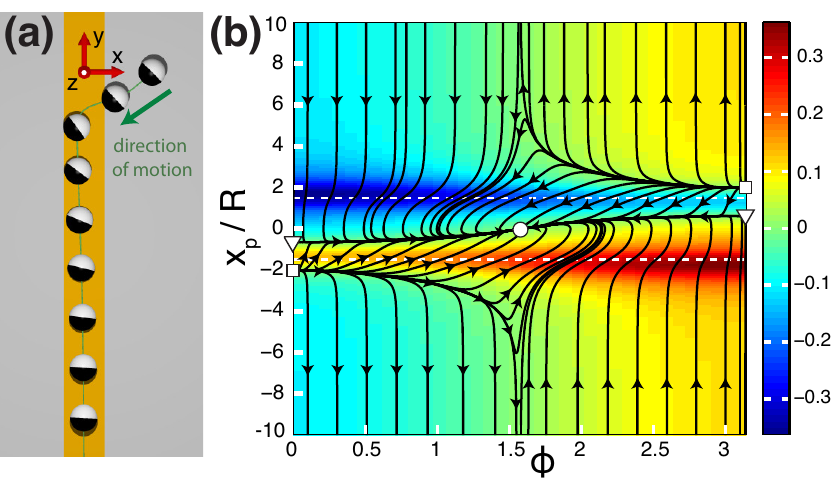}
\caption{\label{fig:figure_3} (a) Trajectory of a catalyst-forward 
($U^{sd}/U_{0}=-0.15$, where $U_{0}\equiv2|b_{w}^{c}|\alpha_{0}/D$) 
particle with $h/R=1.1$ near a chemical stripe of width $2W/R=3$,
 $b_{w}<0$, and $b_{w}/b_{w}^{c}=3$. (b) Phase plane for the particle and the stripe in 
(a). There is an attractor (white circle) at $x_{p}=0$ and $\phi=\pm90^{\circ}$ 
(we recall the mirror symmetry with respect to $\phi=180^{\circ}$). Additionally, there are saddle points (white triangles) and 
unstable fixed points (white squares). The background color encodes 
$U_{x}^{tot}/U_{0}$. (a) and (b) were calculated with the BEM.}
\end{figure}
\begin{figure}[!ht]
\includegraphics[width=8.cm]{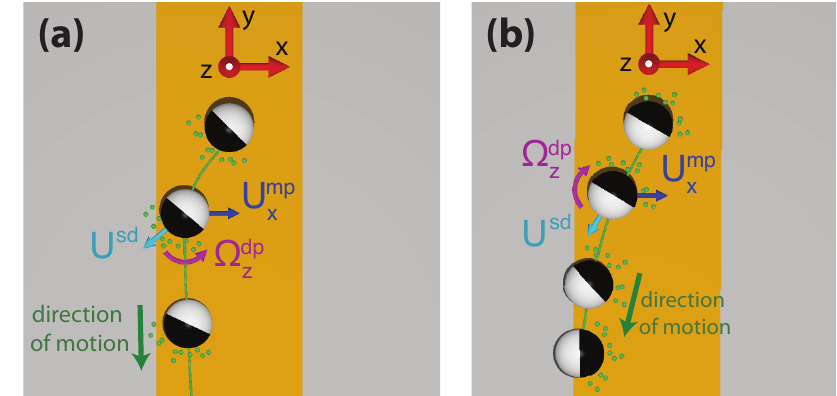}
\caption{\label{fig:stability_mechanism} Illustration of the mechanisms that (a) stabilize the stripe-bound state of a catalyst-forward particle and (b) destabilize the stripe-bound state of an inert-forward particle.
}
\end{figure}

\textit{Chemical stripe.}-- \tm{Next, we consider whether a particle can follow a 
stripe of width $2W$} which has $b=b_{w}^{c}$ (\textit{c} for \textit{c}enter), with $b=b_{w}$ on the rest 
of the substrate. The lengthy expressions which follow from integration are 
given in Sec. IV.C of the SM \cite{Note3}. Analytical and BEM calculations again show good 
agreement. As shown in Fig. \ref{fig:figure_3}, a {catalyst-forward} 
swimmer can follow a stripe: it is attracted to the center and aligns its axis 
parallel to the edges of the stripe ($\phi=\pm90^\circ$). The attraction to the stripe center 
is driven by $U_{x}^{mp}$. At the center, the contributions to $U_{x}^{mp}$ from 
the two edges cancel. In order to understand the stability of the 
alignment $\phi=\pm90^\circ$, we consider a small perturbation  $\delta\phi$. The particle 
starts moving towards one of the edges because for $\delta\phi\neq0$ one 
has $U^{sd}_{x}\neq0$. The edge drives rotation of the cap into the stripe, 
dampening $\delta\phi$ for a catalyst-forward swimmer 
(Fig \ref{fig:stability_mechanism}(a)). For an inert-forward particle, edge 
induced rotation enhances $\delta\phi$, and, for small $U^{sd}$, the 
particle docks (Fig. \ref{fig:stability_mechanism}(b)). A stripe can capture even 
very fast catalyst-forward swimmers: for $|U^{sd}|/U_{0}\gg1$, the basin of 
attraction decreases in size, but the attractor persists (\will{Fig. 5} in the SM \cite{Note3}). 

\textit{Conclusions.}--{Using analytical arguments, supported by detailed numerical calculations, we predict that the motion of a catalytic Janus particle can be controlled via chemical patterning of a confining wall. The pattern ``shapes'' the chemi-osmotic flows on the wall induced by particle's activity. In turn, these flows drive translation and rotation of the particle with respect to the pattern-defined direction. The interplay of chemi-osmosis and self-diffusiophoresis induces two classes of behavior which depend, generically, on whether self-diffusiophoretic motion  is catalyst- or inert-forward. Catalyst-forward particles can stably follow a chemical stripe, while inert-forward particles can dock at the chemical step between two substrate materials. }


\will{Throughout this study, we have focused on particles which maintain a constant height above a wall and an orientation within the plane of the wall.} \will{In two respects, in future research this quasi-2D motion could be realized without the use of external forces.} \tm{First, we note that for two given surfaces that are uniformly composed of distinct materials, the parameters of a Janus swimmer may be} \will{chosen} \tm{such that it will have surface-bound ``sliding'' states \cite{uspal15}, i.e., steady $h$ and $\mathbf{\hat{d}}$, at both surfaces. Such a ``designed'' Janus swimmer might self-adjust to approximate quasi-2D motion near a wall patterned with both materials. 
Secondly, instead of spherical swimmers, one may use heavy rod-like particles 
which, in order to lower their center of gravity, would settle to the in-plane orientation near the bottom wall of a containing vessel. Steric interactions with the wall would prevent significant rocking of the particles. Preliminary calculations confirm that rod-like active particles indeed exhibit a similar phenomenology.} 

\acknowledgments
We thank C. Pozrikidis for making the \texttt{BEMLIB} 
library freely available \cite{pozrikidis02}. W.E.U., M.T., and M.N.P. acknowledge financial support from 
the German Science Foundation (DFG), grant no. TA 959/1-1.

\bibliography{wall_slip}

\newpage



\section*{Supplementary material}

# I. CALCULATION OF CHEMI-OSMOTIC CONTRIBUTION TO THE PARTICLE VELOCITY

Here we detail how to obtain Eq. (1) in the main text from the Lorentz reciprocal theorem. According to this theorem, the fluid stresses $(\boldsymbol{\sigma}, \boldsymbol{\sigma}')$ and the velocity fields $(\mathbf{u}, \mathbf{u}')$ of two solutions to the Stokes equation within the same domain are related by a surface integral over the fluid domain boundaries:

$$\int \mathbf{u} \cdot \boldsymbol{\sigma}' \cdot \mathbf{n} \, dS = \int \mathbf{u}' \cdot \boldsymbol{\sigma} \cdot \mathbf{n} \, dS \;. \tag{1}$$

The "unprimed" solution is for the problem, described in the main text, of determining the chemi-osmotic contributions $\mathbf{U}^{ws}$ and $\boldsymbol{\Omega}^{ws}$ to the motion of the particle. For this problem, the boundary conditions are: $\mathbf{u} = \mathbf{v}_s$ at the planar wall, $\mathbf{u} = \mathbf{U}^{ws} + \boldsymbol{\Omega}^{ws} \times (\mathbf{x} - \mathbf{x}_0)$ at the particle surface, and $\mathbf{u} = 0$ at infinity. Note that here $\mathbf{v}_s$ has already been determined from the solution of the Laplace equation; thus it is a known quantity.

Since there are six unknowns $\mathbf{V} = (\mathbf{U}^{ws}, \boldsymbol{\Omega}^{ws})$, six "primed" problems are required. These problems are indexed by $j = 1, \ldots, 6$. For $j = 1, 2, 3$, we consider a particle subject to an external force with amplitude $F'^{ext}$ in $\hat{x}$, $\hat{y}$, or $\hat{z}$, respectively. For $j = 4, 5, 6$, the particle is subject to an external torque with amplitude $\tau'^{ext}$ in $\hat{x}$, $\hat{y}$, or $\hat{z}$, respectively. For each of the cases $j = 1, \ldots, 6$ we impose that the motion is subject to no-slip boundary conditions, i.e., $\mathbf{u}'^{(j)} = 0$ at the planar wall, and that the fluid is quiescent far away from the particle, i.e., $\mathbf{u}'^{(j)} = 0$ at infinity. At the particle surface, there is a no-slip condition which implies $\mathbf{u}'^{(j)} = \mathbf{U}'^{(j)} + \boldsymbol{\Omega}'^{(j)} \times (\mathbf{x} - \mathbf{x}_0)$. Here, $\mathbf{U}'^{(j)}$ and $\boldsymbol{\Omega}'^{(j)}$ are the unknown translational and angular velocities of the particle driven by the external force or external torque in problem $j$.

We apply the Lorentz theorem (Eq. (1)) to each of the six pairs obtained by combining the unprimed problem with the subproblem $j$. At the planar wall, $\mathbf{u}'^{(j)}$ vanishes, and far away from the particle, it decays at least as fast as $1/r$. Therefore, concerning the integral over the whole boundary of the fluid domain, only the part over the surface of the particle contributes. The velocity field $\mathbf{u}$ decays at least as fast as $1/r$, and therefore only the surface of the particle and the wall contribute to the integral involving $\mathbf{u}$. This leads to the following

set of equations:

$$\int_{z=0} \mathbf{u} \cdot \boldsymbol{\sigma}'^{(j)} \cdot \mathbf{n} \, dS + \int_{|\mathbf{r}|=R} \mathbf{u} \cdot \boldsymbol{\sigma}'^{(j)} \cdot \mathbf{n} \, dS = \int_{|\mathbf{r}|=R} \mathbf{u}'^{(j)} \cdot \boldsymbol{\sigma} \cdot \mathbf{n} \, dS, \quad j = 1, \ldots, 6, \qquad (2)$$

where we have split the lhs into two integrals, and $\mathbf{r} \equiv \mathbf{x} - \mathbf{x_0}$.

We now show that the rhs is identically zero. To this end we insert the boundary conditions $\mathbf{u}'^{(j)}$ and obtain

$$\text{rhs} = \int_{|\mathbf{r}|=R} \left( \mathbf{U}'^{(j)} + \boldsymbol{\Omega}'^{(j)} \times \mathbf{r} \right) \cdot \boldsymbol{\sigma} \cdot \mathbf{n} \, dS.$$

We consider the two terms on the rhs in turn. For the translational term, we have

$$\int_{|\mathbf{r}|=R} \mathbf{U}'^{(j)} \cdot \boldsymbol{\sigma} \cdot \mathbf{n} \, dS = \mathbf{U}'^{(j)} \cdot \int_{|\mathbf{r}|=R} \boldsymbol{\sigma} \cdot \mathbf{n} \, dS = \mathbf{U}'^{(j)} \cdot \mathbf{F}, \qquad (3)$$

where $\mathbf{F} = \int_{|\mathbf{r}|=R} \boldsymbol{\sigma} \cdot \mathbf{n} \, dS$ is, by definition [1], the force exerted by the fluid on the self-propelled particle (plus its thin boundary layer of thickness $\delta$). Since this is the only force acting on the particle, and since the self-propelled particle is force-free ($\mathbf{F} = 0$), this term vanishes. For the rotational term, we have

$$\int_{|\mathbf{r}|=R} \boldsymbol{\Omega}'^{(j)} \times \mathbf{r} \cdot \boldsymbol{\sigma} \cdot \mathbf{n} \, dS = \boldsymbol{\Omega}'^{(j)} \cdot \int_{|\mathbf{r}|=R} \mathbf{r} \times \boldsymbol{\sigma} \cdot \mathbf{n} \, dS = \boldsymbol{\Omega}'^{(j)} \cdot \boldsymbol{\tau}. \qquad (4)$$

Using the vector identity $(\mathbf{a} \times \mathbf{b}) \cdot \mathbf{c} = \mathbf{a} \cdot (\mathbf{b} \times \mathbf{c})$ we have rearranged the integrand and identified the last integral with the torque exerted by the fluid on the self-propelled particle (plus its thin boundary layer of thickness $\delta$) [1]. Since the self-propelled particle is torque-free ($\boldsymbol{\tau} = 0$), the rotational term vanishes, too. Therefore, the entire right hand side of Eq. (2) is zero, which leads to

$$\int_{z=0} \mathbf{u} \cdot \boldsymbol{\sigma}'^{(j)} \cdot \mathbf{n} \, dS + \int_{|\mathbf{r}|=R} \mathbf{u} \cdot \boldsymbol{\sigma}'^{(j)} \cdot \mathbf{n} \, dS = 0, \quad j = 1, \ldots, 6. \qquad (5)$$

(Note that, at this point, the unknown quantities $\mathbf{U}'^{(j)}$ and $\boldsymbol{\Omega}'^{(j)}$ have dropped out of these

equations.) We now substitute the boundary conditions in the lhs and rearrange:

$$\int_{|\mathbf{r}|=R} \mathbf{U}^{ws} \cdot \boldsymbol{\sigma}'^{(j)} \cdot \mathbf{n}\, dS + \int_{|\mathbf{r}|=R} \boldsymbol{\Omega}^{ws} \times \mathbf{r} \cdot \boldsymbol{\sigma}'^{(j)} \cdot \mathbf{n}\, dS = -\int_{z=0} \mathbf{v}_s \cdot \boldsymbol{\sigma}'^{(j)} \cdot \mathbf{n}\, dS. \quad (6)$$

Due to manipulations similar to the above ones, we obtain

$$\mathbf{U}^{ws} \cdot \mathbf{F}'^{(j)} + \boldsymbol{\Omega}^{ws} \cdot \boldsymbol{\tau}'^{(j)} = -\int_{z=0} \mathbf{v}_s \cdot \boldsymbol{\sigma}'^{(j)} \cdot \mathbf{n}\, dS\ ,\ j=1,\ldots,6\ , \quad (7)$$

where $\mathbf{F}'^{(j)}$ and $\boldsymbol{\tau}'^{(j)}$ are the force and torque, respectively, *from the fluid on the particle* in subproblem $j$.

In addition to the generalized velocity vector $\mathbf{V}$ introduced above, we also define a generalized force $\boldsymbol{\mathfrak{F}}'^{(j)} \equiv (\mathbf{F}'^{(j)}, \boldsymbol{\tau}'^{(j)})$. In each subproblem $j$, the component $\mathfrak{F}'^{(j)}_j$, such as $F'^{(1)}_x$ or $\tau'^{(4)}_x$, must exactly cancel the imposed force or torque because the motion of the particle is overdamped ($Re \ll 1$). It is therefore known *a priori*. (For instance, $F'^{(1)}_x = -F'^{ext}$ and $\tau'^{(4)}_x = -\tau'^{ext}$.) The other, off-diagonal components, such as $\tau'^{(1)}_y$, are unknown prior to finding the solution of subproblem $j$. However, the off-diagonal terms are significant only if the particle is very close to the wall, i.e., when $h/R \approx 1$ [2]. We therefore neglect the off-diagonal terms, and obtain

$$\mathfrak{F}'^{(j)}_j V_j \approx -\int \mathbf{v}_s \cdot \boldsymbol{\sigma}'^{(j)} \cdot \mathbf{n}\, dS,\ j=1,...,6 \quad (8)$$

where $V_j$ are components of the generalized velocity $\mathbf{V}$. (Note that the left hand side of Eq. (8) is not a sum over $j$.) Due to the linearity of the Stokes equation, $\boldsymbol{\sigma}'^{(j)}$ contains as a prefactor either $F'^{ext}$ or $\tau'^{ext}$, i.e., $-\mathfrak{F}'^{(j)}_j$. Therefore the (arbitrary) amplitudes $F'^{ext}$ or $\tau'^{ext}$ drop out of the problem. In order to avoid a clumsy notation, in the following we shall denote $\boldsymbol{\sigma}'^{(j)}/\mathfrak{F}'^{(j)}_j$ as $\boldsymbol{\sigma}'^{(j)}$.

## II. MULTIPOLE EXPANSION FOR SOLUTE CONCENTRATION

In order to perform the integral in Eq. (8) analytically, two expressions are needed: an expression for $\mathbf{v}_s$, and an expression for $\boldsymbol{\sigma}'^{(j)}$. For both quantities, we use "point-particle" approximations in order to obtain analytically tractable expressions. In this section, we obtain expressions for the surface concentration gradient $\nabla_{||}c(\mathbf{x}_s)$, recalling from the main text that $\mathbf{v}_s(\mathbf{x}_s) = -b_w(\mathbf{x}_s)\nabla_{||}c(\mathbf{x}_s)$.

For a spherical particle with axisymmetric catalyst coverage in *free space* (i.e., far from bounding surfaces), the solute number density can be expanded in Legendre polynomials [3]:

$$c^{fs}(r,\theta') = c_\infty + \frac{R}{D}\sum_{l=0}^{\infty} \frac{\alpha_l}{l+1}\left(\frac{R}{r}\right)^{l+1} P_l(\cos(\theta')), \qquad (9)$$

where $\alpha_l$ are the multipole coefficients of the surface activity (areal density per time) $\alpha(\theta') = \sum_{l=0}^{\infty} \alpha_l P_l(\cos(\theta'))$, $\mathbf{r}$ is the vector from the center of the particle to an observation point, $r = |\mathbf{r}|$, and $\theta'$ is the angle between $\mathbf{r}$ and the vector $\hat{\mathbf{d}}$ oriented along the axis of symmetry of the particle. As defined in the main text, $\hat{\mathbf{d}}$ points from the catalytic cap of the particle to the inert region. For any specification $\alpha(\theta')$ of the particle activity, the coefficients $\alpha_l$ are easily calculated. For instance, for the constant-flux model of activity presented in the main text, one has the monopole coefficient $\alpha_0 = \kappa/2$ and the dipole coefficient $\alpha_1 = -3\kappa/4$. (The sign of $\alpha_1$ is negative due to our choice of direction for $\hat{\mathbf{d}}$.)

Now we consider an active particle near a planar wall in a configuration in which $\hat{\mathbf{d}}$ is parallel to the wall (which occupies the $xy$ plane); thus $\hat{\mathbf{d}} = (\cos(\phi),\sin(\phi),0)$. (The conditions under which such configurations can be realized are discussed in the main text.) In order to obtain approximate analytical expressions for the surface gradient $\nabla_{||}c(\mathbf{x}_s)$, we make two approximations: (i) We truncate the multipole expansion for the activity of the particle, and consider only the monopole and dipole terms. (ii) In order to model the effect of confinement of the solute field by the wall, we place mirror images of the monopole and the dipole below the wall. We neglect additional reflections of these two images across the particle surface. *A posteriori* the validity of these assumptions is checked via comparisons of the theoretically predicted dynamical behavior of the particle with that obtained from BEM numerical solutions of the full problem.

We therefore write $c(\mathbf{x}_s) \approx c^{mp}(\mathbf{x}_s) + c^{dp}(\mathbf{x}_s)$. Here, $c(\mathbf{x}_s)$ is the sum of two terms: the field

due to a point source (*mono*pole) of the number density located at the center $\mathbf{x}_0 = (x_p, y_p, h)$ of the particle plus the field due to an *i*mage point source of the number density located at $\mathbf{x}_I = (x_p, y_p, -h)$. The second term, $c^{dp}(\mathbf{x}_s)$, is a contribution from a *di*pole and its image. The real and image dipoles are likewise located at $\mathbf{x}_0$ and the image point $\mathbf{x}_I$, respectively, and both have the strength $\mathbf{p} = -|\alpha_1|\hat{\mathbf{d}}$.

For the monopole term, we obtain

$$c^{mp}(\mathbf{x}_s) = \frac{2\alpha_0 R^2}{Dr}, \tag{10}$$

$$\nabla_{||} c^{mp}(\mathbf{x}_s) = -\frac{2\alpha_0 R^2 \mathbf{r}_s}{Dr^3}. \tag{11}$$

Here, $\mathbf{x}_s = (x, y, 0)$ denotes a point on the wall, $\mathbf{r}_s = (x - x_p, y - y_p, 0)$, and $r = \sqrt{(x - x_p)^2 + (y - y_p)^2 + h^2}$. For the dipole term, we obtain

$$c^{dp}(\mathbf{x}_s) = -\frac{|\alpha_1| R^3}{D} \frac{\hat{\mathbf{d}} \cdot \mathbf{r}_s}{r^3}, \tag{12}$$

$$\nabla_{||} c^{dp}(\mathbf{x}_s) = -\frac{|\alpha_1| R^3}{Dr^3} \left(\mathbb{1} - \frac{3\mathbf{r}_s \mathbf{r}_s}{r^2}\right) \cdot \hat{\mathbf{d}}. \tag{13}$$

Both $c^{mp}(\mathbf{x}_s)$ and $c^{dp}(\mathbf{x}_s)$ satisfy the no-flux condition $\hat{n} \cdot \nabla c = 0$ on the planar wall. However, our approximation for $c(\mathbf{x}_s)$ does not account for the finite size of the particle. The main effect of the finite size of the particle is that of strongly confining the solute between the wall and the particle surface when the ratio $h/R$ is approaching 1, creating there a region of high number density of solute. It has been shown that, at $\mathcal{O}((R/h)^2)$, accounting for the finite size of the particle requires the introduction of an image dipole located at the center of the particle and pointing towards the wall [4]. Since this term accounts for a number density distribution *at the wall* having the same *in plane* rotational symmetry as the monopole, it does not introduce additional features, but quantitatively enhances the strength of $c^{mp}$. Accounting for the finite size at $\mathcal{O}((R/h)^3)$ requires the introduction of an image dipole at the particle center which is oriented parallel to the wall [4]. It likewise has only a quantitative effect of a somewhat increased strength of $c^{dp}$. Hence, while our approximation for $c^{mp}(\mathbf{x})$ is not exact to $\mathcal{O}((R/h)^2)$, it captures the main symmetries of the system and thus the physical phenomena of interest.

## III. APPROXIMATION FOR THE SHEAR STRESS AT THE WALL

We seek an analytical expression for the shear stress $\boldsymbol{\sigma}'^{(j)}$ to be substituted into Eq. (8). To this end we obtain a "point-particle" approximation as follows.

For the "primed" problems $j = 1, 2, 3$, we replace the particle by a point force (Stokeslet) pointing into the directions $\hat{x}$, $\hat{y}$, and $\hat{z}$, respectively, located at $\mathbf{x}_0 = (x_p, y_p, h)$ (i.e., the center of the particle) above a planar wall located at $z = 0$. The fluid satisfies incompressibility and the Stokes equations. As shown by Blake [5], the governing equations and the no-slip condition on the wall can be satisfied by locating a system of images, which consists of a Stokeslet, a force-dipole, and a source-doublet (see Ref. [5]), at the point $\mathbf{x}_I = (x_p, y_p, -h)$. For the sake of clarity, we introduce the mapping $(1, 2, 3, 4, 5, 6) \overset{s}{\to} (x, y, z, x, y, z)$ connecting the "primed" problem index and the corresponding direction of the unit force or torque (see also below), as well as the shorthand notation $s_j := s(j)$ (i.e., $s_1 := s(1) = x$, $s_5 := s(5) = y$, etc.). The corresponding fluid velocity at an observation point $\mathbf{x} = (x, y, z)$ is given by

$$u'^{(j)}_i = \frac{1}{8\pi\eta}\left[\left(\frac{1}{r} - \frac{1}{X}\right)\delta_{is_j} + \frac{r_i r_{s_j}}{r^3} - \frac{X_i X_{s_j}}{X^3} + 2h(\delta_{s_j\alpha}\delta_{\alpha l} - \delta_{s_j z}\delta_{zl})\frac{\partial}{\partial X_l}\left[\frac{hX_i}{X^3} - \left(\frac{\delta_{iz}}{X} + \frac{X_i X_z}{X^3}\right)\right]\right], \quad (14)$$

where $i \in \{x, y, z\}$, $\alpha \in \{x, y\}$, $\mathbf{r} \equiv \mathbf{x} - \mathbf{x}_0$, $\mathbf{X} \equiv \mathbf{x} - \mathbf{x}_I$, $r \equiv |\mathbf{r}|$, $X \equiv |\mathbf{X}|$, and the Einstein convention of summation over repeated indices is used (here and in the following). Note that the first product of Kroneker delta symbols vanishes when either $s_j$ or $l$ take the value $z$, while the second such product contributes only when both $s_j$ and $l$ take the value $z$. Therefore, the index $l$ in the partial derivative with respect to $X_l$ is taken to be $x$, $y$, or $z$. The pressure is given by

$$P'^{(j)} = \frac{1}{4\pi}\left[\frac{r_{s_j}}{r^3} - \frac{X_{s_j}}{X^3} - 2h(\delta_{s_j\alpha}\delta_{\alpha l} - \delta_{s_j z}\delta_{zl})\frac{\partial}{\partial X_l}\left(\frac{X_z}{X^3}\right)\right]. \quad (15)$$

(Although pressure is a scalar quantity, $s_j$ appears in the expression for the pressure because its functional form depends on the direction of the point force.) From the velocity and pressure it follows that the stress tensor

$$\boldsymbol{\sigma}' = -P'\mathbb{1} + \eta(\nabla\mathbf{u}' + \nabla\mathbf{u}'^T) \quad (16)$$

in the fluid is given by the following expression:

$$\sigma'^{(j)}_{ik} = \frac{3}{4\pi\eta}\left[\frac{r_i r_{s_j} r_k}{r^5} - \frac{X_i X_{s_j} X_k}{X^5} - 2h(\delta_{s_j\alpha}\delta_{\alpha l} - \delta_{s_j z}\delta_{zl})\left(-\frac{h}{X^5}\delta_{ik}X_l \right.\right.$$
$$\left.\left. + \frac{z}{X^5}(X_i\delta_{lk} + X_k\delta_{il}) + \frac{X_i X_k}{X^5}\delta_{zl} - \frac{5z X_i X_l X_k}{X^7}\right)\right] \quad (17)$$

This expression approximately recovers the stress from a sphere dragged by an external point force in the presence of a wall at $z = 0$, although it neglects the finite size of the sphere (represented, for hydrodynamics, by a no-slip condition on the surface of the sphere.)

For substitution into the reciprocal theorem, we are interested in the quantities $\sigma'^{(j)}_{iz}|_{z=0}$, where $i \in \{x, y\}$. These quantities are components of the shear stress evaluated at the wall. We obtain

$$\sigma'^{(1)}_{iz}|_{z=0} = -\frac{3h}{2\pi}\frac{r_i r_x}{r^5}, \quad (18)$$

$$\sigma'^{(2)}_{iz}|_{z=0} = -\frac{3h}{2\pi}\frac{r_i r_y}{r^5}, \quad (19)$$

and

$$\sigma'^{(3)}_{iz}|_{z=0} = \frac{3h^2}{2\pi}\frac{r_i}{r^5}. \quad (20)$$

Now we turn to the "primed problems" $j = 4, 5, 6$. For these three problems, we consider a point torque oriented into the directions $\hat{x}$, $\hat{y}$, and $\hat{z}$, respectively, located at $\mathbf{x}_0$. For a point torque above a planar wall at $z = 0$, Blake found [5], via the method of images, that the velocity field is

$$u'^{(j)}_i = \frac{1}{8\pi\eta}\left[\epsilon_{is_j k}\left(\frac{r_k}{r^3} - \frac{X_k}{X^3}\right) + 2h\epsilon_{ks_j z}\left(\frac{\delta_{ik}}{X^3} - \frac{3X_i X_k}{X^5}\right) + 6\epsilon_{ks_j z}\frac{X_i X_k X_z}{X^5}\right], \quad (21)$$

where $\epsilon_{\alpha\beta\gamma}$ denotes the Levi-Civita symbol (with the convention that its indices are interpreted as $x \to 1$, $y \to 2$, and $z \to 3$), and, as mentioned above, summation over repeated indices is employed here and the following. The pressure is [5]

$$P'^{(j)} = -4\eta\frac{\partial}{\partial X_k}\left(\frac{\epsilon_{ks_j z}X_z}{X^3}\right). \quad (22)$$

By using the definition (Eq. (16)) for calculating the stress tensor $\boldsymbol{\sigma}'^{(j)}$, we obtain the corresponding $iz$ components ($i \in \{x, y\}$) evaluated at the wall, which are needed for substitution in the reciprocal theorem:

$$\sigma'^{(4)}_{iz}|_{z=0} = -\frac{3}{4\pi}\frac{\delta_{iy}h^2 - r_i r_y}{r^5}, \quad (23)$$

$$\sigma'^{(5)}_{iz}|_{z=0} = -\frac{3}{4\pi}\frac{\delta_{ix}h^2 - r_i r_x}{r^5}, \tag{24}$$

and

$$\sigma'^{(6)}_{iz}|_{z=0} = \frac{3h}{4\pi}\frac{\epsilon_{izk}r_k}{r^5}. \tag{25}$$

# IV. CHEMI-OSMOTIC FLOW CONTRIBUTIONS TO THE PARTICLE VELOCITY: ANALYTICAL EXPRESSIONS

Since $\hat{\mathbf{d}}$ is assumed to remain parallel to the planar wall at $z = 0$, only the components $j = 1, 2$, and $6$ of the generalized velocity $\mathbf{V}$ are of interest (which are equal to $U_x, U_y$, and $\Omega_z$, respectively; see Sec. I). We calculate the individual contributions of the monopole ($^{mp}$) and dipole ($^{dp}$) number density terms (see Eqs. (11) and (13)), respectively, to these velocity components. This is carried out for the cases of a homogeneous substrate, a chemical step, and a chemical stripe. In each case, the components of the stress tensor $\boldsymbol{\sigma}'^{(j)}|_{z=0}$ required for the calculation of $U_x$ ($j = 1$), $U_y$ ($j = 2$), and $\Omega_z$ ($j = 6$) are provided by Eqs. (18), (19), and (25), respectively.

## A. Uniform substrate

For a uniform substrate with surface mobility $b_w$, Eq. (8) yields

$$V_j = b_w \int_{-\infty}^{-\infty} dx \int_{-\infty}^{\infty} dy \, \left( \nabla_{||} c \cdot \boldsymbol{\sigma}'^{(j)} \cdot \mathbf{n} \right) |_{z=0} . \tag{26}$$

The monopole contributions are obtained by replacing $\nabla_{||} c$ in Eq. (26) with the corresponding expression in Eq. (11). The result is

$$U_x^{mp} = U_y^{mp} = 0 \,, \Omega_z^{mp} = 0 \,. \tag{27}$$

This is expected because in the plane of the wall the number density distribution due to a monopole above the wall is radially symmetric around $(x_p, y_p)$, and therefore the flow it induces cannot drive translations parallel to the plane or in-plane rotations of $\hat{\mathbf{d}}$. Similarly, the dipole contributions are obtained by replacing $\nabla_{||} c$ in Eq. (26) with the corresponding expression in Eq. (13). After performing the resulting integrals, we obtain

$$U_x^{dp} = -\frac{b_w R^3 |\alpha_1|}{16 D h^3} \cos(\phi) \tag{28}$$

$$U_y^{dp} = -\frac{b_w R^3 |\alpha_1|}{16 D h^3} \sin(\phi) \tag{29}$$

and
$$\Omega_z^{dp} = 0, \tag{30}$$

where $\phi$ is the angle between $\hat{\mathbf{d}}$ and $\hat{x}$. We have therefore obtained that, above a uniform substrate, the dipolar contribution drives chemi-osmotic "surfing," i.e., translation in the $\hat{\mathbf{d}}$ direction: $\mathbf{U}^{dp} = -\frac{b_w R^3 |\alpha_1|}{16 D h^3} \hat{\mathbf{d}}$.

We note that "surfing" can change the inert-forward or catalyst-forward character of motion in $\hat{\mathbf{d}}$ near a surface from that observed in the bulk. For such a change to occur, one must have $|b_w| \gg |b_p|$, where $b(\mathbf{x}_s) \sim b_p$ at the particle surface. This special case requires that the particle and substrate materials have strongly different strengths of interaction with the solute.

### B. Chemical step

We now consider a wall with a chemical step between two materials, such that $b(\mathbf{x}_s) = b_w^l$ for $x < 0$ and $b(\mathbf{x}_s) = b_w^r$ for $x > 0$. Accordingly, Eq. (8) is evaluated piecewise:

$$V_j = b_w^l \int_{-\infty}^{0} dx \int_{-\infty}^{\infty} dy \left( \nabla_{\|} c \cdot \boldsymbol{\sigma}'^{(j)} \cdot \mathbf{n} \right) + b_w^r \int_{0}^{\infty} dx \int_{-\infty}^{\infty} dy \left( \nabla_{\|} c \cdot \boldsymbol{\sigma}'^{(j)} \cdot \mathbf{n} \right). \tag{31}$$

Following the line of the derivations in the previous subsection, after straightforward but cumbersome algebra we obtain the contributions from the monopole term,

$$U_x^{mp} = \frac{3 h R^2 \alpha_0}{16 D} \left[ \frac{(b_w^r - b_w^l)(h^2 + 2 x_p^2)}{(h^2 + x_p^2)^{5/2}} \right] \tag{32}$$

and
$$U_y^{mp} = 0, \Omega_z^{mp} = 0, \tag{33}$$

and from the dipole term:

$$U_x^{dp} = \frac{|\alpha_1| R^3}{256 D h^3} \left[ -8(b_w^r + b_w^l) + (b_w^l - b_w^r) \frac{(-25 h^4 + 28 h^2 x_p^2 + 8 x_p^4) x_p^3}{(h^2 + x_p^2)^{7/2}} \right] \cos(\phi), \tag{34}$$

$$U_y^{dp} = \frac{|\alpha_1| R^3}{256 D h^3} \left[ -8(b_w^r + b_w^l) + (b_w^l - b_w^r) \frac{(8 x_p^4 + 20 h^2 x_p^2 + 3 h^4) x_p}{(h^2 + x_p^2)^{5/2}} \right] \sin(\phi), \tag{35}$$

and
$$\Omega_z^{dp} = -\frac{3 h R^3 |\alpha_1|}{64 D} \frac{(b_w^l - b_w^r)}{(h^2 + x_p^2)^{5/2}} \sin(\phi). \tag{36}$$

Note that, as expected, the results do not depend on $y_p$, because the system exhibits translational symmetry along the $y$-direction.

Interestingly, $\Omega_z^{dp}$ is mirror symmetric with respect to $x_p = 0$ (see also Fig. 2(b) of the main text.) This is a generic feature of $\Omega_z^{ws}$ (which also holds for any contribution from higher multipole moments), and can be understood as follows. Since the Stokes equation is linear, the contributions of the left and right hand sides of the substrate surface to $\Omega_z^{ws} = 0$ can be calculated separately and linearly superposed. (For instance, in order to calculate the contribution of the left hand side, one takes $b(\mathbf{x}_s) = b_w^l$ for $x < 0$ and $b(\mathbf{x}_s) = 0$ for $x > 0$ and determines the corresponding particle velocity.) Accordingly, $\Omega_z^{ws}$ can be expressed as a linear combination of two functions: $\Omega_z^{ws} = b_w^l g^l(x_p) + b_w^r g^r(x_p)$. Since $\Omega_z^{ws} = 0$ for a uniform substrate ($b_w^l = b_w^r = b_w$), the two functions are linked as $g^l(x_p) = -g^r(x_p)$. Additionally, mirror symmetry about $x = 0$ requires that the two functions are related by $g^l(x_p) = -g^r(-x_p)$ (see below and Fig. 1); when combined with $g^l(x_p) = -g^r(x_p)$, it yields that $g^r(x_p)$ is even: $g^r(x_p) = g^r(-x_p)$. The implication of the mirror symmetry can be understood as follows. Fig. 1(a) illustrates the contribution $\Omega_z^l = b_w g^l(x_p)$ of the left hand side (shaded area) to the angular velocity of a particle at point $x_p$, where we have taken $b_w^l = b_w$. The rotation can be either clockwise or counterclockwise; here, clockwise is chosen for illustration. Applying a mirror transformation about $x = 0$ leads to the situation shown in Fig. 1(b). The sign of the angular velocity has changed (here, from clockwise to counterclockwise.) However, Fig. 1(b) can also be interpreted as depicting the contribution $\Omega_z^r = b_w g^r(-x_p)$ to the angular velocity of a particle at $-x_p$ from the right hand side of the substrate surface for $b_w^r = b_w$. Accordingly, one obtains the relation $g^l(x_p) = -g^r(-x_p)$ noted above. This implies $\Omega_z^{ws} = b_w^l g^l(x_p) + b_w^r g^r(x_p) = -b_w^l g^r(-x_p) + b_w^r g^r(x_p) = (b_w^r - b_w^l) g^r(x_p)$, where $g^r(x_p)$ is an even function of $x_p$.

The threshold condition for docking at the step is derived as follows. We recall from the main text that for an inert-forward particle ($U^{sd} > 0$) docking occurs for $x_p \lesssim 0$ and $\phi = 0°$ if $b_w^l > b_w^r$, and for $x_p \gtrsim 0$ and $\phi = 180°$ if $b_w^l < b_w^r$. In order to obtain the characteristic scale $U_{x0}^{mp}$ discussed in the main text, in Eq. (32) we set $x_p = 0$:

$$U_{x0}^{mp} = \frac{3\alpha_0 R^2 (b_w^r - b_w^l)}{16 D h^2}. \tag{37}$$

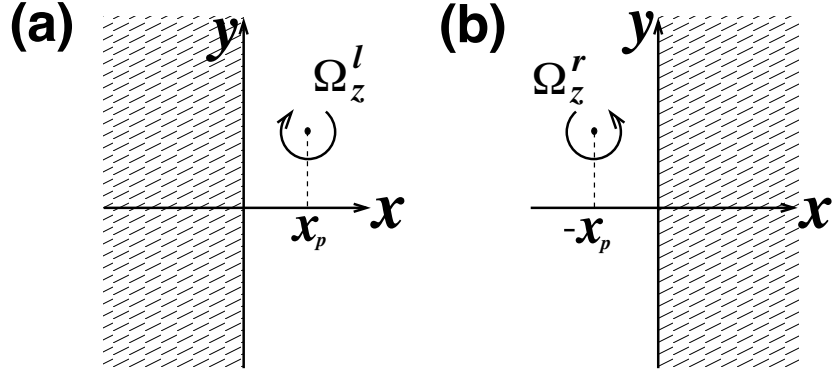

FIG. 1: (a) Schematic illustration of the contribution $\Omega_z^l = b_w g^l(x_p)$ of the left hand side of the substrate surface $z = 0$ (shaded area) to the angular velocity of a particle at $x_p$ for $b_w^l = b_w$. As discussed in the text, this contribution is calculated by taking $b_w^r = 0$ on the right hand side of the substrate surface (white area). $\Omega_z^l$ can be either clockwise or counterclockwise; here, clockwise is chosen for illustration. (b) Upon applying a mirror transformation about $x = 0$ to (a), one finds that the resulting contribution has changed sign. However, panel (b) also illustrates the contribution $\Omega_z^r = b_w g^r(-x_p)$ to the angular velocity of a particle at $-x_p$ from the right hand side of the substrate surface (shaded area) for $b_w^r = b_w$. Accordingly, one has $g^l(x_p) = -g^r(-x_p)$.

In order to obtain $U_{x0}^{dp}$, in Eq. (34) we set $x_p = 0$:

$$U_{x0}^{dp} = -\frac{|\alpha_1| R^3 (b_w^r + b_w^l)}{32 D h^3} \cos(\phi) \tag{38}$$

Since in the docking configuration one has $U_x^{tot} := U_x^{sd} + U_x^{mp} + U_x^{dp} = 0$, where $U_x^{sd} = U^{sd} \cos(\phi)$, we therefore obtain the threshold condition:

$$-U_c^{sd} \cos(\phi) = \frac{3\alpha_0 R^2 (b_w^r - b_w^l)}{16 D h^2} - \frac{|\alpha_1| R^3 (b_w^r + b_w^l)}{32 D h^3} \cos(\phi). \tag{39}$$

If $b_w^l > b_w^r$, one has $\phi = 0°$ at the docking configuration, whereas for $b_w^l < b_w^r$, one has $\phi = 180°$ at the docking configuration. Accordingly, we obtain:

$$-U_c^{sd} = \frac{3\alpha_0 R^2 (b_w^r - b_w^l)}{16 D h^2} - \frac{|\alpha_1| R^3 (b_w^r + b_w^l)}{32 D h^3}, \quad b_w^l > b_w^r. \tag{40}$$

$$U_c^{sd} = \frac{3\alpha_0 R^2 (b_w^r - b_w^l)}{16 D h^2} + \frac{|\alpha_1| R^3 (b_w^r + b_w^l)}{32 D h^3}, \quad b_w^l < b_w^r. \tag{41}$$

Equations. (40) and (41) can be combined into

$$U_c^{sd} = \frac{3\alpha_0 R^2 |b_w^l - b_w^r|}{16Dh^2} + \frac{|\alpha_1| R^3 (b_w^r + b_w^l)}{32Dh^3}, \qquad (42)$$

which is the form used in the main text.

## C. Stripe

We now consider a substrate with a chemical stripe of width $2W$. We take $b(\mathbf{x}_s) = b_w^l$ for $x < -W$, $b(\mathbf{x}_s) = b_w^c$ for $-W < x < W$, and $b(\mathbf{x}_s) = b_w^r$ for $x > W$. In the main text, we discuss the special case $b_w^r = b_w^l \equiv b_w$. However, for reasons of generality, in the following equations we allow the left and right hand sides of the stripe to differ. The integral consists of three pieces:

$$\begin{aligned} V_j = b_w^l \int_{-\infty}^{-W} dx \int_{-\infty}^{\infty} dy \, (\nabla_{||} c \cdot \boldsymbol{\sigma}'^{(j)} \cdot \mathbf{n}) + b_w^c \int_{-W}^{W} dx \int_{-\infty}^{\infty} dy \, (\nabla_{||} c \cdot \boldsymbol{\sigma}'^{(j)} \cdot \mathbf{n}) + \\ b_w^r \int_{W}^{\infty} dx \int_{-\infty}^{\infty} dy \, (\nabla_{||} c \cdot \boldsymbol{\sigma}'^{(j)} \cdot \mathbf{n}) . \end{aligned} \qquad (43)$$

Along the lines of Subsec. IV.B, we obtain

$$U_x^{mp} = \frac{3hR^2 \alpha_0}{16D} \left[ \frac{(b_w^c - b_w^l)(h^2 + 2(x_p + W)^2)}{(h^2 + (x_p + W)^2)^{5/2}} + \frac{(b_w^r - b_w^c)(h^2 + 2(x_p - W)^2)}{(h^2 + (x_p - W)^2)^{5/2}} \right], \qquad (44)$$

$$\begin{aligned} U_x^{dp} = -\frac{|\alpha_1| R^3}{256 D h^3} \bigg[ &-8(b_w^r + b_w^l) \\ &+ (b_w^c - b_w^r)\frac{(-25h^4 + 28h^2(x_p - W)^2 + 8(x_p - W)^4)(x_p - W)^3}{(h^2 + (x_p - W)^2)^{7/2}} \\ &+ (b_w^l - b_w^c)\frac{(-25h^4 + 28h^2(x_p + W)^2 + 8(x_p + W)^4)(x_p + W)^3}{(h^2 + (x_p + W)^2)^{7/2}} \bigg] \cos(\phi), \end{aligned} \qquad (45)$$

$$\begin{aligned} U_y^{dp} = \frac{|\alpha_1| R^3}{256 D h^3} \bigg[ &-8(b_w^r + b_w^l) \\ &+ (b_w^c - b_w^r)\frac{(8(x_p - W)^4 + 20h^2(x_p - W)^2 + 3h^4)(x_p - W)}{(h^2 + (x_p - W)^2)^{5/2}} \\ &+ (b_w^l - b_w^c)\frac{(8(x_p + W)^4 + 20h^2(x_p + W)^2 + 3h^4)(x_p + W)}{(h^2 + (x_p + W)^2)^{5/2}} \bigg] \sin(\phi), \end{aligned} \qquad (46)$$

and
$$\Omega_z^{dp} = -\frac{3hR^3|\alpha_1|}{64D}\left[\frac{(b_w^c - b_w^r)}{(h^2 + (x_p - W)^2)^{5/2}} + \frac{(b_w^l - b_w^c)}{(h^2 + (x_p + W)^2)^{5/2}}\right]\sin(\phi). \quad (47)$$

We note that, as in the case of a chemical step, there is no dependence on $y_p$ due to the translational invariance of the system along the $y$-direction.

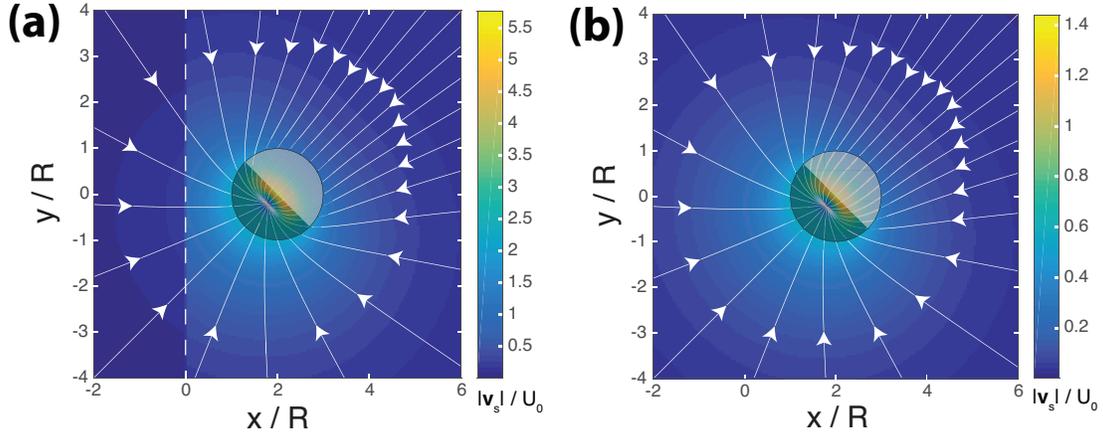

FIG. 2: (a) Streamlines on a substrate with a chemical step. Here, one has $b_w^r/b_w^l = 4$, $b_w^l < 0$, $h/R = 1.1$, $x_p = 2$, and $\phi = 45°$. The color indicates the surface flow strength $|\mathbf{v}(\mathbf{x}_s)|/U_0$. (b) Streamlines of chemi-osmotic flow on a homogeneous substrate for $h/R = 1.1$ and $b_w < 0$, calculated within BEM.

## V. CHEMI-OSMOTIC SURFACE FLOWS

Some intuition for the physics behind the chemi-osmotic contributions to the particle velocity can be developed from a more detailed consideration of the surface flows. The surface flows entrain flows in the bulk solution, which couple to the particle. Near the wall, the streamlines of the bulk flows approximately follow the streamlines of the surface flows. For a Janus particle above a homogeneous substrate, the chemi-osmotic flows on the substrate must be mirror symmetric, owing to the symmetry of the particle and of the wall configuration ($\hat{\mathbf{d}}$ parallel to the wall). The mirror plane is defined such that it contains $\hat{\mathbf{d}}$ (which is along the axis of symmetry of the particle) and the surface normal $\hat{\mathbf{z}}$. Recalling the restrictions on the particle configuration (fixed $h/R$, and $\hat{\mathbf{d}}$ confined to a plane parallel to the substrate), the only possible contribution of the surface flow to the particle velocity is translational motion along the axis of symmetry i.e, in the $\hat{\mathbf{d}}$ or $-\hat{\mathbf{d}}$. In particular, chemi-osmotic flows cannot rotate the particle around the $\hat{z}$ axis.

In Fig. 2(b), we show streamlines of surface flows for a particle above a uniform substrate, calculated with the BEM. As expected, the streamlines are mirror symmetric with respect to the midplane orthogonal to the substrate. They are directed quasi-radially inward to a point displaced from the particle center towards the cap. In the near-wall region of the bulk

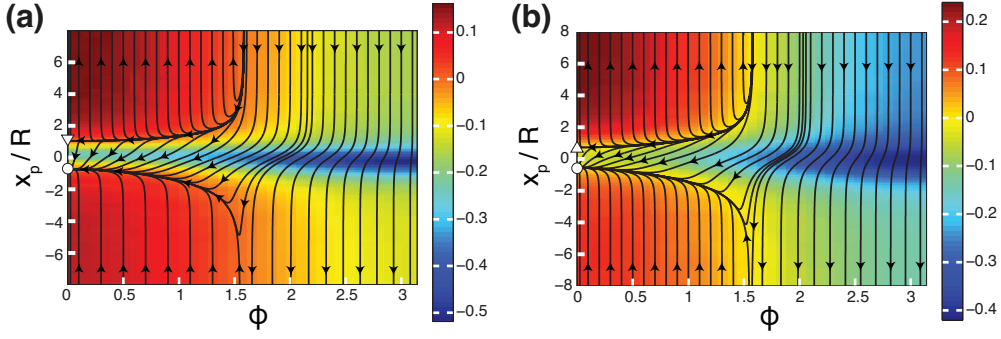

FIG. 3: (a) Phase plane calculated within BEM for a chemical step with $b_w^r/b_w^l = 4$, $b_w^l < 0$ for a particle with $U^{sd}/U_0 = 0.1$ and $h/R = 1.1$, where $U_0 \equiv 2|b_w^l|\alpha_0/D$. There is an attractor (white circle) at $\phi = 0°$ and $x_p/R = -0.69$ and an unstable fixed point (white triangle). The background color gives $U_x^{tot}/U_0$. This phase plane is also shown in Fig. 2 of the main text; it is reproduced here for comparison with (b). (b) Phase plane calculated by using the approximate expressions given by Eqs. (32), (34), (35) (36), and $U^{sd}/U_0 = 0.1$ (same as in (a)). There is an attractor (white circle) at $\phi = 0°$ and $x_p/R = -0.59$, as well as an unstable fixed point (white triangle).

fluid (not shown), the flows approximately follow the streamlines of the surface flows, except near the point where the surface streamlines converge. Here, incompressibility requires the bulk flows to lift off from the substrate. Although the streamlines shown correspond to an *exact* solution obtained within BEM, they exhibit a characteristic "monopole plus dipole" pattern. This result strongly suggests that truncating the particle activity at the dipole level is a reliable approximation.

In Fig. 2(a), we show streamlines of surface flows for a particle above a chemical step $(b_w^r/b_w^l = 4, b_w^l < 0)$. Strikingly, the streamlines of the surface flow are identical to those for a patterned substrate. This is because the patterning scales the *magnitude* of surface flow $|\mathbf{v}_s|$ at each point, and hence does not affect the local *direction*. Therefore, the "monopole plus dipole" pattern of the streamlines is, in this sense, *universal*.

It is interesting to note that the directions of rotation and translation can also be inferred reliably from thermal equilibrium arguments, i.e., taking the direction of motion to be such as to reduce the free energy of the solute. Similarly, for self-diffusiophoresis of a colloid, reasoning in terms of thermal equilibrium renders the correct direction of motion, but not the correct dependence on the material parameters of the system [6].

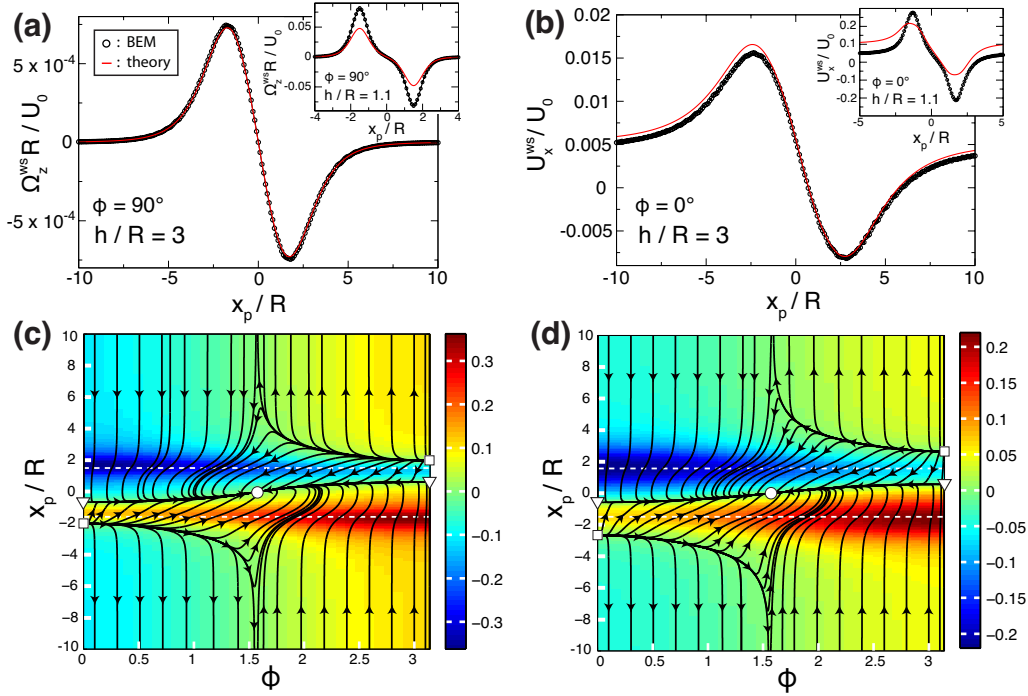

FIG. 4: (a) Angular velocity as a function of $x_p$ for a particle oriented parallel ($\phi = 90°$) to the stripe in Fig. 3 of the main text ($2W/R = 3$, $b_w/b_w^c = 3$, and $b_w < 0$), here with $U_0 \equiv 2|b_w^c|\alpha_0/D$. In the main panel one has $h/R = 3$ and in the inset $h/R = 1.1$. Black circles have been obtained with the BEM, and the red curve stems from Eq. 3 of the main text. (b) Chemi-osmotic contribution to $U_x$ as a function of $x_p$ for a particle with $\phi = 0°$ for the same stripe. (c) Phase plane calculated with BEM for the same stripe as in (a) and (b) and $h/R = 1.1$ with $U^{sd}/U_0 = -0.15$. There is an attractor (white circle) at $\phi = \pi/2$ and $x_p = 0$. Additionally, there are two saddle points (white triangles) and two unstable fixed points (white squares). The background color gives $U_x^{tot}/U_0$. This phase plane is also shown in Fig. 3(b) of the main text; it is reproduced here for comparison with (d). (d) Phase plane calculated by using the approximate expressions given by Eqs. (44), (45), (47), and $U^{sd}/U_0 = -0.15$ (same as in (c)). The other parameters have the same values as in (a), (b), and (c).

## VI. TRAJECTORIES NEAR A CHEMICAL STEP OR STRIPE

In Figs. 3(a) and (b), we show phase planes calculated for the chemical step considered in Fig. 2 of the main text. Here, the particle has $h/R = 1.1$ and $U^{sd}/U_0 = 0.1$. Figure 3(a) was obtained by using BEM, and Fig. 3(b) by using the analytical expressions; they are almost identical.

Now we consider a chemical stripe. In Fig. 4(a) and (b), we compare the expressions derived in Sec. IV C with full calculations obtained by using BEM. The parameters characterizing the stripe are the same as in Fig. 3 of the main text. As in the case of a chemical

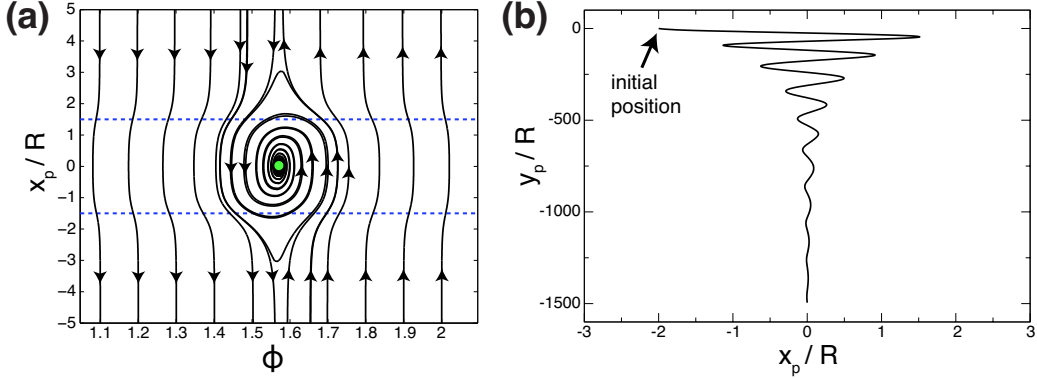

FIG. 5: (a) Phase plane for a very fast catalyst-forward particle ($U^{sd}/U_0 = -10$, $h/R = 1.1$) near the stripe, as discussed in Fig. 4 and also in Fig. 3 of the main text ($2W/R = 3$, $b_w/b_w^c = 3$, and $b_w < 0$, with $U_0 \equiv 2|b_w^c|\alpha_0/D$). Remarkably, there is still an attractor (green circle) at $x_p = 0$, $\phi = \pi/2$, but the size of the basin of attraction is considerably reduced compared with the one in Fig. 4(c). (b) Trajectory for the same fast particle and the same stripe as in (a), with initial conditions $x_p/R = -2$, $y_p = 0$, and $\phi = \pi/2$. The particle exhibits a decaying oscillatory motion about the center of the stripe. The results in (a) and (b) have been calculated with the BEM.

step, we find excellent quantitative agreement between the approximate analytical expressions and the BEM for $h/R = 3$. For $h/R = 1.1$, the approximations capture the qualitative features of the BEM data, but quantitative discrepancies are evident. In Figs. 4(c) and (d), we show phase planes calculated for $h/R = 1.1$ by using BEM and by using the analytical expressions, respectively. As with the chemical step, it can be seen that the analytical predictions recover almost quantitatively the topology of the phase plane (i.e., the stable and unstable fixed points, the saddle points, the shape and the extent of the basin of attraction).

Remarkably, the stripe-following state is stable even for very fast catalyst-forward particles, as shown in Fig. 5. Figure 5(a) shows a section of the phase plane calculated with BEM for a particle with $U^{sd}/U_0 = -10$ and $h/R = 1.1$ near the stripe from Fig. 4 (and also Fig. 3 of the main text.) There is still an attractor, although its basin of attraction no longer spans the whole range of particle orientations $\phi$ (compare Fig. 4(c)). Interestingly, the particle configuration $(x_p, \phi)$ approaches the attractor through decaying oscillations (see the spiraling phase trajectory). Figure 5(b) shows a trajectory in the $xy$ plane, i.e., in real space, with the initial conditions $x_p/R = -2$, $y_p = 0$, and $\phi = \pi/2$. The trajectory exhibits decaying oscillations around the stripe center $x_p = 0$. The decay length of these oscillations is large when measured by the distance traveled by the particle in the $\hat{y}$-direction; within a rough estimate, the particle is captured by the center when $y_p/R = -500$. The survival of

the stripe-following attractor for very large $|U^{sd}|/U_0$ suggests that substrate materials which interact only very weakly with the solute, can still, if arranged into a stripe pattern, guide the motion of catalyst-forward swimmers.

---

26



\end{document}